\documentclass[article]{aa}
\usepackage{graphicx}
\usepackage{lscape}

\def\msol{\hbox{\kern 0.20em $M_\odot$}}

\newcommand{\lsol}{\hbox{\kern 0.20em $L_\odot$}}
\newcommand{\g}{\hbox{\kern 0.20em g}}
\newcommand{\gmu}{\hbox{\kern 0.20em g$^{-1}$}}
\newcommand{\kg}{\hbox{\kern 0.20em kg}}
\newcommand{\pc}{\hbox{\kern 0.20em pc}}
\newcommand{\mum}{\hbox{\kern 0.20em $\mu$m}}
\newcommand{\mumd}{\hbox{\kern 0.20em $\mu$m$^{-2}$}}
\newcommand{\cm}{\hbox{\kern 0.20em cm}}
\newcommand{\m}{\hbox{\kern 0.20em m}}
\newcommand{\km}{\hbox{\kern 0.20em km}}
\newcommand{\nm}{\hbox{\kern 0.20em nm}}
\newcommand{\s}{\hbox{\kern 0.20em s}}
\newcommand{\h}{\hbox{\kern 0.20em h}}
\newcommand{\smu}{\hbox{\kern 0.20em s$^{-1}$}}
\newcommand{\smd}{\hbox{\kern 0.20em s$^{-2}$}}
\newcommand{\an}{\hbox{\kern 0.20em an}}
\newcommand{\anmu}{\hbox{\kern 0.20em an$^{-1}$}}
\newcommand{\yr}{\hbox{\kern 0.20em yr}}
\newcommand{\yrmu}{\hbox{\kern 0.20em yr$^{-1}$}}
\newcommand{\Myr}{\hbox{\kern 0.20em Myr}}
\newcommand{\Mymu}{\hbox{\kern 0.20em Myr$^{-1}$}}
\newcommand{\K}{\hbox{\kern 0.20em K}}
\newcommand{\pcmu}{\hbox{\kern 0.20em pc$^{-1}$}}
\newcommand{\pcmd}{\hbox{\kern 0.20em pc$^{-2}$}}
\newcommand{\pcmt}{\hbox{\kern 0.20em pc$^{-3}$}}
\newcommand{\kms}{\hbox{\kern 0.20em km\kern 0.20em s$^{-1}$}}
\newcommand{\kmpd}{\hbox{\kern 0.20em km$^{2}$}}
\newcommand{\kpc}{\hbox{\kern 0.20em kpc}}
\newcommand{\cms}{\hbox{\kern 0.20em cm\kern 0.20em s$^{-1}$}}
\newcommand{\erg}{\hbox{\kern 0.20em erg}}
\newcommand{\ergs}{\hbox{\kern 0.20em erg}}
\newcommand{\cmpd}{\hbox{\kern 0.20em cm$^2$}}
\newcommand{\cmmd}{\hbox{\kern 0.20em cm$^{-2}$}}
\newcommand{\cmms}{\hbox{\kern 0.20em cm$^{-6}$}}
\newcommand{\cmpt}{\hbox{\kern 0.20em cm$^3$}}
\newcommand{\cmmt}{\hbox{\kern 0.20em cm$^{-3}$}}
\newcommand{\mpd}{\hbox{\kern 0.20em m$^2$}}
\newcommand{\mmd}{\hbox{\kern 0.20em m$^{-2}$}}
\newcommand{\mpt}{\hbox{\kern 0.20em m$^3$}}
\newcommand{\mmt}{\hbox{\kern 0.20em m$^{-3}$}}
\newcommand{\mujy}{\hbox{\kern 0.20em $\mu$Jy}}
\newcommand{\mjy}{\hbox{\kern 0.20em mJy}}
\newcommand{\Mj}{\hbox{\kern 0.20em MJy}}
\newcommand{\jy}{\hbox{\kern 0.20em Jy}}
\newcommand{\ghz}{\hbox{\kern 0.20em GHz}}
\newcommand{\G}{\hbox{\kern 0.20em G}}
\newcommand{\muG}{\hbox{\kern 0.20em $\mu$G}}

\newcommand{\htwo}{\hbox{H${}_2$}}

\newcommand{\hto}{\hbox{H${}_2$O}}

\newcommand{\jtwotoone}{\hbox{$J=2\rightarrow 1$}}
\newcommand{\jthreetotwo}{\hbox{$J=3\rightarrow 2$}}

\newcommand{\jfourtothree}{\hbox{$J=4\rightarrow 3$}}
\newcommand{\jeighttoseven}{\hbox{$J=8\rightarrow 7$}}

\newcommand{\water}{\hbox{H$_{2}$O}}

\newcommand{\pwater}{\hbox{p-H$_{2}$O {}$3_{13}-2_{20}$}}

\begin{document}
\title{Shocked water in the Cep\,E protostellar outflow}
\author{
B.~Lefloch\inst{1,2} \and J.~Cernicharo\inst{2} \and S.~Pacheco\inst{1} \and
C.~Ceccarelli\inst{1}}

\offprints{lefloch@obs.ujf-grenoble.fr}

\institute{UJF-Grenoble 1 / CNRS-INSU, Institut de Plan\'etologie et d'Astrophysique de Grenoble (IPAG) UMR 5274, Grenoble, F-38041, France; e-mail: lefloch@obs.ujf-grenoble.fr
 \and Laboratorio de Astrof\'{\i}sica Molecular, Departamento de
Astrof\'{\i}sica, Centro de Astrobiologia, INTA, Ctra de Torrej\'on a Ajalvir, km 4,
E-28850 Torrej\'on de Ardoz, E-28850 Madrid, Spain
 }

\date{Received~: 3 December 2010; Accepted~: 10 January 2011}
\abstract {Previous far-infrared observations at low-angular resolution have
reported the presence of water associated with low-velocity outflow shocks and
protostellar envelopes.}{We want to elucidate the origin of water emission in
protostellar systems.}{The outflow driven by the intermediate-mass class 0
protostar Cep\,E is among the most luminous outflows detected so far. Using the
IRAM 30m telescope,
we searched for and detected the \pwater\ line
emission at $183\ghz$ in the Cep\,E star-forming core. The emission arises from
high-velocity gas close to the protostar, which is unresolved in the main beam of
the telescope. Complementary observations at $2\arcsec$ resolution with the Plateau
de Bure interferometer
helped establish the origin of the emission detected and the physical conditions
in the emitting gas. The water line profile and its spatial distribution are
very similar to those of SiO. We find that the \water\ emission arises from
warm ($\sim 200\K$), dense ($(1-2)\times 10^6\cmmt$) gas, and its abundance is
enhanced by one to two orders of magnitude with respect to the protostellar
envelope.}{We detect water emission in strong shocks from the high-velocity jet
at 1000~AU from the protostar. Despite the large beam size of the
telescope, such emission should be detectable with Herschel.}{}

\keywords{Stars: formation - ISM: individual (Cep\,E) - ISM: jets and outflows - ISM: molecules}

\maketitle

\section{Introduction}

Water is a key molecule not only for oxygen chemistry in the interstellar gas,
but also in the dynamical evolution of star-forming regions, because it is one of
the main gas coolants (Kaufman \& Neufeld, 1996; Bergin et al. 1998). Whereas
its abundance is very low in dark clouds, prestellar cores, and starless regions
(from less than $10^{-8}$ to a few $10^{-7}$; Bergin \& Snell, 2002; Moneti et
al. 2001), it can be greatly enhanced in star-forming regions, through the
sputtering of frozen water from grain mantles and through high-temperature,
sensitive reactions in the gas phase (Elitzur \& de Jong, 1978; Ceccarelli et
al. 1996; Kaufmann \& Neufeld 1996). It has been proposed that such processes
are driven by the heating of the nascent stars (hot core/corino region) or by
shocks either in the accretion region in the inner protostellar core or in the
outflow carrying away the angular momentum of the envelope. Previous
observations of water lines obtained with the {\em Infrared Space Observatory}
 have substantially confirmed these theoretical
expectations (see  e.g. Giannini et al. 2001; Ceccarelli et al. 2000; van Dishoeck 2004
for a review). However, many details are still missing and the relative
efficiency of these mechanisms is not known, so that the origin of the water
emission in protostellar regions remains a puzzle to be solved. One obstacle is
indeed the relatively low spatial resolution of these observations.

Most of the \water\ lines cannot be observed from ground because of strong
atmospheric absorption. The para-\water\ line $3_{13}-2_{20}$ at 183.3~GHz is
an exception since it can be enhanced by the masing effect even in regions of moderate
excitations.
Using the IRAM 30m telescope, Cernicharo et al. (1990,1994)  detected
large-scale \pwater\ line emission  in giant molecular clouds and star-forming
regions and they showed that this line can be used to investigate the spatial
distribution of \hto\ in protostellar environments at an angular resolution
comparable to the size of the protostellar cores ($14\arcsec$ at the IRAM 30m
telescope). Observing the low-mass protostellar outflows HH~7-11 and L1448,
Cernicharo et al. (1996) detected water line emission arising from shocks
associated with the Herbig-Haro objects in the outflows.

In this Letter, we report on observations of the \pwater\ line in the
star-forming core Cep\,E with the IRAM 30m telescope\footnote{Based on
observations carried out with the IRAM 30m Telescope.
IRAM is supported by INSU/CNRS (France), MPG (Germany), and IGN (Spain)}. Located at 730pc, it is a
relatively nearby intermediate-mass Class~0 source of $100 \lsol$ and a mass of
$18\msol$, intensively studied at mm and IR wavelengths (Moro-Martin et al.
2001; Noriega-Crespo et al. 2004; Lefloch et al. 1996, hereafter L96). The
protostar drives an exceptionally powerful outflow, whose southern lobe is terminated
by the Herbig-Haro object HH~377. Far-infrared observations at
low spectral and angular resolution with ISO have revealed large amounts of water,
which Giannini et al. (2001) interpret as arising from low-velocity shocks,
whose far-infrared luminosity amounts to $\rm L\sim 3\lsol$.

Observations of the millimeter lines of SiO observations reveal the presence of
strong shocks along the outflow (L96). Because it is usually undetected in the
cold, quiescent molecular gas, SiO is a particularly good tracer of shocks
that are strong enough to release refractory elements from grain cores. Recent {\em
Herschel} observations suggest that SiO is also a good tracer of the water-emitting
shock regions (Lefloch et al. 2010; Nisini et al. 2010). We present
observations of the high-excitation SiO \jeighttoseven\ line towards the
protostar.

\section{Observations}

%figure 1
\begin{figure}
\includegraphics[width=0.49\columnwidth]{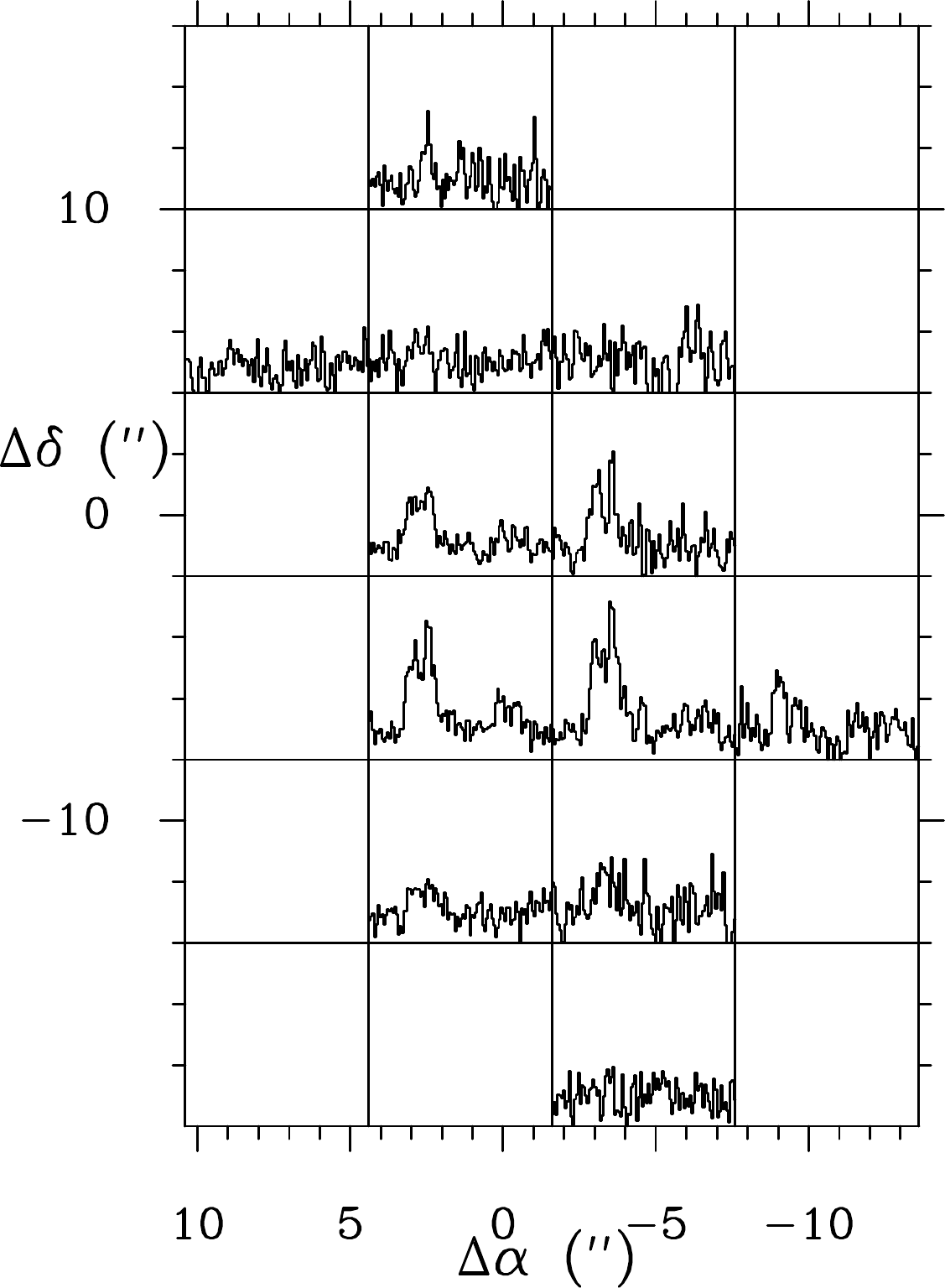}
\includegraphics[width=0.49\columnwidth]{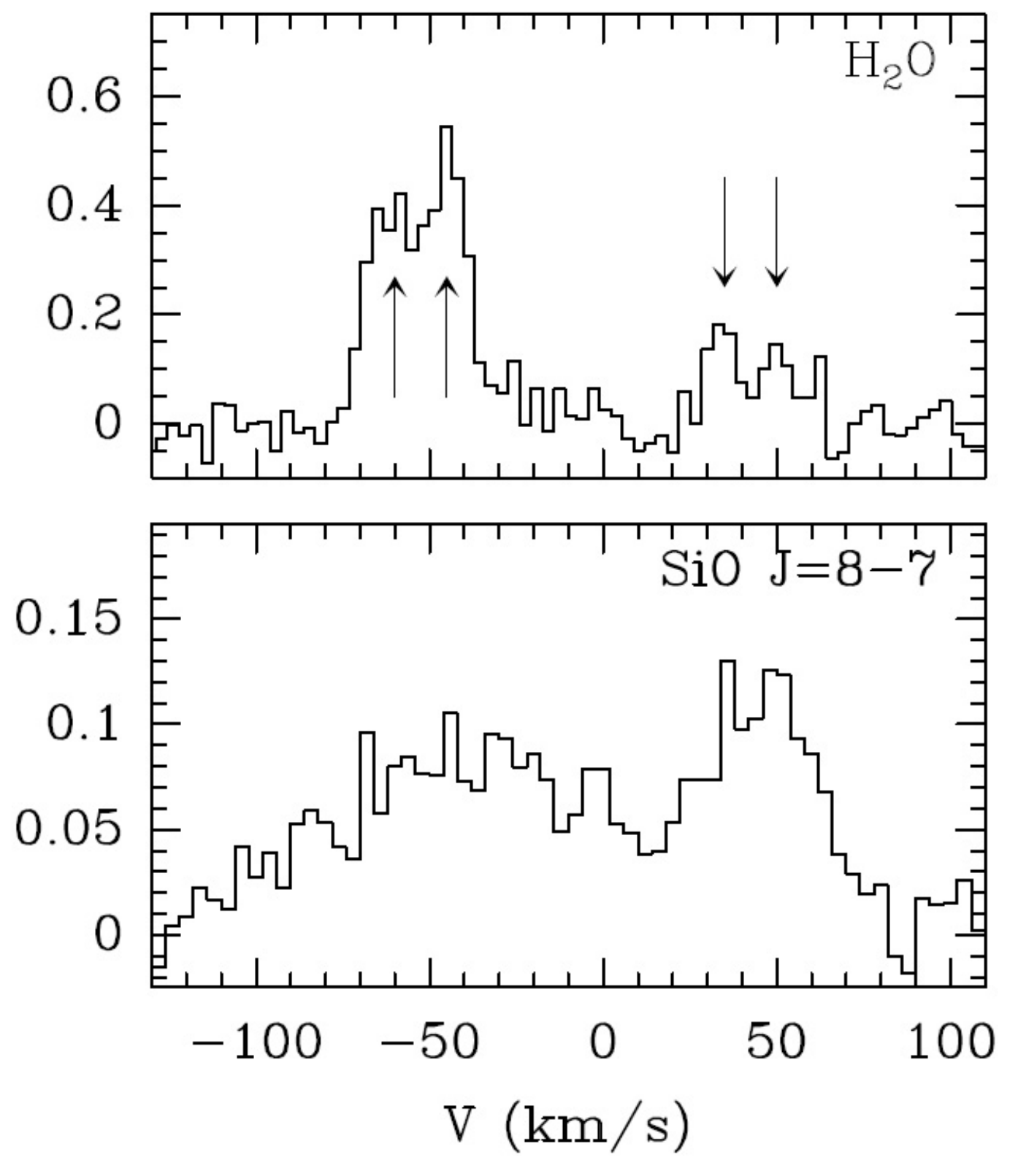}
\caption{{\em (left)}~Map of the \pwater\ emission between $-110$ and $+70\kms$
in the outflow of Cep\,E obtained with the IRAM 30m telescope. Coordinates are
in arcsec offsets with respect to the nominal position of the protostar. {\em
(right)}~\hto\ and SiO \jeighttoseven\ emission detected towards the
protostellar source. The \hto\ emission is averaged over a region of $6\arcsec
\times 6\arcsec$ around the protostar. Intensities are expressed in units of
antenna temperature. The different components of the spectrum are marked with
arrows. } \label{iramradio}
\end{figure}

\subsection{Water observation}

We observed the line of para-\water\ $3_{13}\rightarrow 2_{20}$  at 183.3 GHz
at the IRAM 30m telescope in January 2009, near the transit of Cep\,E
between $61^{\circ}$ and $65^{\circ}$ of elevation.
Both receivers at 2mm were used simultaneously.  Observations were carried
out in wobbler switching mode with a
throw of $180\arcsec$. Filterbanks with 1MHz spectral resolution and an
autocorrelator providing a spectral resolution of 1.25~MHz (2 km/s) were used
as spectrometers.  Weather conditions were very good and stable,
with an atmospheric optical depth $\approx 1.10$ along the line of sight.
Calibration was carefully monitored every 4 minutes between two
integration scans on Cep\,E. Water vapor and system temperatures were found to
remain stable, in the range $1650-1800\K$. Any fluctuation of the sky emissivity
between the on and off positions would result in a poor spectroscopic baseline and
detection of the mesospheric water line, which is not the case in the present
observations. Pointing was checked against nearby quasars every hour and has an
accuracy of $\approx 2\arcsec$. At the frequency of the \hto\ line, the
main-beam efficiency of the telescope is 0.63 and the half-power beamwidth is
$13.2\arcsec$. We did a small oversampled map at $6\arcsec$ sampling of the
southern lobe of the outflow, centered on the nominal position of the protostar
$23^h 03^m 13.0^s$ $+61^{\circ} 42\arcmin 21\arcsec$ (J2000).  The typical rms
is 100mK per velocity interval of $2\kms$.

\subsection{SiO observations}

The SiO \jeighttoseven\ transition at 347.33006~GHz was observed towards the
protostar at the JCMT in August 2005. The weather conditions were good with an
average system temperature $\rm T_{sys}\sim 400\K$. Observations were carried
out with a nutating secondary, adopting a throw of $60\arcsec$. The DBS was
used as spectrometer, providing a spectral resolution of 625kHz. The resolution
was degraded to 3.75 MHz, corresponding to a final velocity resolution of about
$3.2\kms$. At the frequency of the SiO line, the main-beam efficiency of the
telescope is about 0.60 and the half-power beamwidth is $\simeq 14.2\arcsec$,
very similar to the telescope parameters of the p-\water\ observations.
Complementary interferometric observations at $2\arcsec$ resolution with the
Plateau de Bure in the SiO \jtwotoone\ transition provide us with the spatial
distribution of the emission and its excitation conditions. These observations
will be presented and discussed in detail in a forthcoming paper (Lefloch et
al. in prep.).

%figure 2
\begin{figure}
\includegraphics[width=\columnwidth]{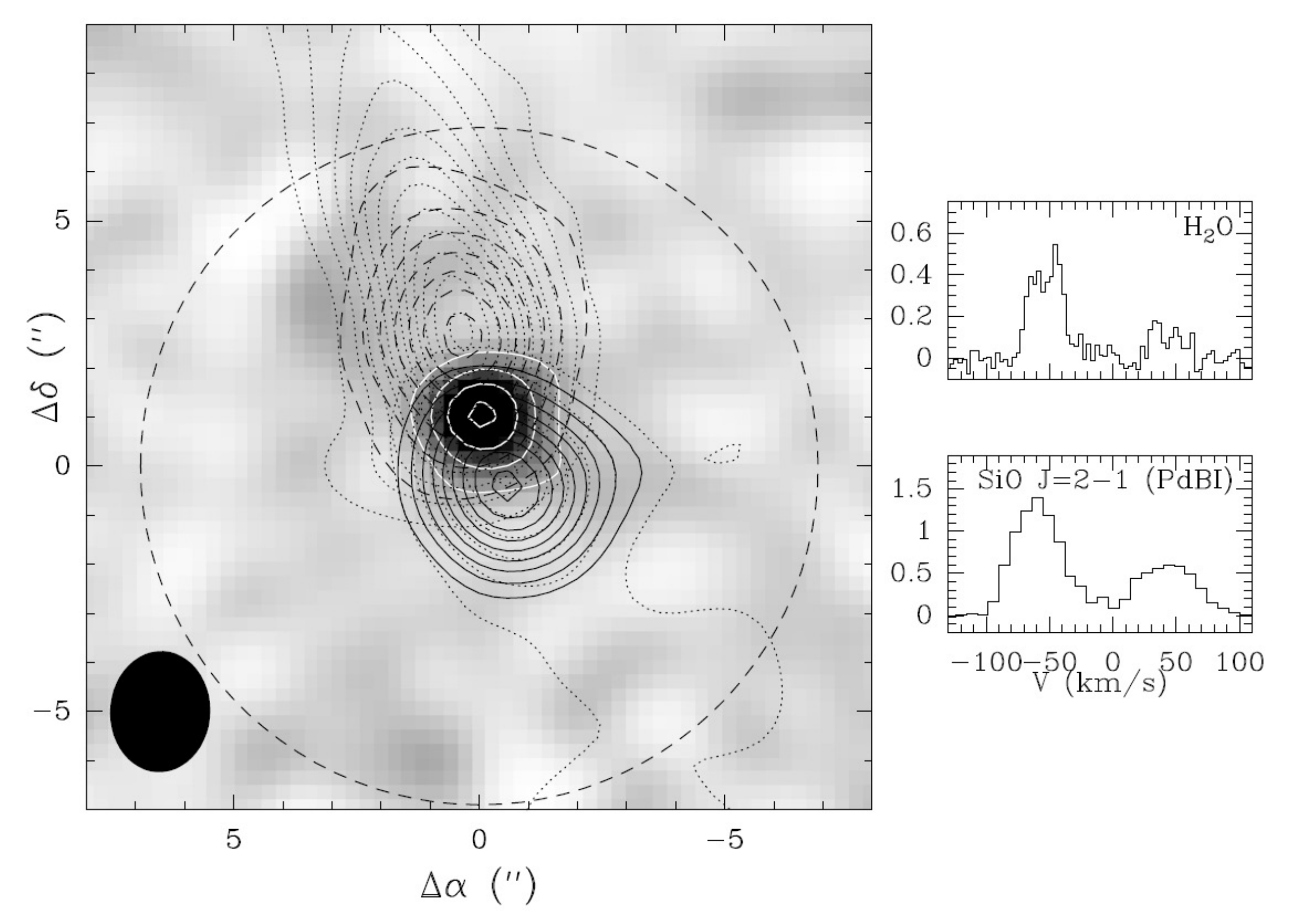}
\caption{({\em left})~Map of the velocity-integrated SiO emission in the Cep\,E
jet obtained at $2\arcsec$ resolution with the Plateau de Bure interferometer
(PdBI; Lefloch et al. in prep.). The SiO $\jtwotoone$ emission integrated
between $+40$ and $+80\kms$ ($-90$ and $-40\kms$) is drawn in solid (dashed)
contours, superposed on the 3~mm continuum emission map (grayscale and white
contours). First contour and contour interval of the jet emission are $10^{-2}$
and $5\times 10^{-3}$ ($10^{-2}$) $\rm Jy/beam$ for the redshifted
(blueshifted) emission. First contour and contour interval of the 3mm continuum
emission is $\rm 2\times 10^{-3}\rm Jy/beam$. The main beam of the IRAM 30m
telescope at the frequency of the \pwater\ line  is marked by a circle
(dashed). The synthesized beam of the PdBI ($2.4\arcsec \times 2.0\arcsec$) is
indicated as a filled ellipse in the bottom right corner. ({\em
right})~Comparison of the PdBI SiO $\jtwotoone$ and the average $\pwater$
spectra towards the protostar.  Intensities are expressed in units of antenna
temperature. } \label{iramradio}
\end{figure}

\section{Results and discussion}

\subsection{Water emission in Cep\,E}

The map of the \pwater\ emission in Cep\,E is displayed in the left panel of
Fig.~1. Emission is detected only towards the protostellar core, in a region
unresolved by the main beam of the IRAM 30m telescope. In particular, we do not
detect any \water\ emission towards HH77, the southern bowshock at offset
position ($-5\arcsec$, $-15\arcsec$), associated with bright emission in the
lines of SiO and $\htwo$ (L96). To improve the signal-to-noise ratio
of the data, we averaged the emission of the water line over the inner core, in
the range $\Delta\alpha=$ [$1\arcsec ; -5\arcsec$] $\Delta\delta=$ [$1\arcsec ;
-5\arcsec$]. The averaged spectrum is displayed in the right panel of Fig.~1, and
the final rms noise is 30mK ($T_A^{*}$) per interval of $2\kms$.

Two bright components ($\sim 1\K$) with broad linewidths ($\sim 10\kms$) are
detected at about $v_{lsr}\approx -65$ and $-45\kms$ in the direction of the
protostar. Two other, weaker, \water\ components are detected at approximately
symmetric (redshifted) velocities $v_{lsr}\approx +34$ and $+50\kms$. We do not
detect any emission at the velocity of the protostellar envelope ($v_{lsr}=
-10.9\kms$; L96), hence excluding any contribution from the hot corino to the
water emission. The presence of both blue- and redshifted components in the
water line spectrum indicates that the emission arises from the central
protostellar region. This is because there is barely any overlap in the plane
of the sky between the blue and redshifted wings of the Cep\,E outflow, except
in the central protostellar region as a consequence of the limited angular
resolution of the observations (see Figs.~1-2; also L96).

The spectrum  of the SiO \jeighttoseven\ emission  in the direction of the
protostar is displayed in Fig.~1 and looks very similar to the \pwater\ line
spectrum. Resemblance between both emissions is even more striking when
comparing the water spectrum with the \jtwotoone\ emission observed
with the PdBI (Fig.~2). This
suggests that the emission of both tracers arises from the same region.
Observations of the \jtwotoone\ transition with the PdBI of the central
protostellar regions (Lefloch et al. in prep) show that the SiO emission arises
from two compact clumps, in the blue- and redshifted gas, respectively,  located
at $\approx 1.5\arcsec$ from the protostar. They have a typical size of
$2.5\arcsec$ ($2.3\times 10^{16}\cm$), and a transverse size of $1\arcsec$
(Fig.~2). Their similar physical properties (size, velocity, distance) suggest
they are related to an episodic ejection phenomenon. In what follows, we assume that
both the \water\ and the SiO emissions have the same size.

\subsection{Physical conditions}

The gas kinetic temperature $\rm T_k$ in the redshifted, high-velocity jet  was
previously determined by Hatchell et al. (1999) from JCMT observations of the
CO \jfourtothree\ and \jtwotoone\ lines. Unfortunately, their analysis is
biased by a highly overestimated value of the main-beam efficiency $\eta_{MB}$
of the \jfourtothree\ observations, poorly known at that time, leading to an
underestimate of $T_k$ ($\approx 50\K$). Based on their observations
and adopting a standard value consistent with the subsequent determinations
from point-like
calibrators\footnote{http://www.jach.hawaii.edu/JCMT/spectral$\_$line/Standards/},
$\eta_{MB}= 0.35$, we obtain the jet kinetic temperature and column density,
$T_k \simeq 120\K$ and $\rm N(CO)\simeq 2.5\times 10^{17}\cmmd$, respectively
(see also L96). Adopting a standard CO abundance in the gas, we then derive the
source-averaged gas column-density $\rm N(\htwo)\simeq 2.5\times 10^{21}\cmmd$. We adopt
these values of $\rm T_k$ and N(CO) for both water-emitting clumps at blue- and
redshifted velocities.

The gas density and column density in the water-emitting clumps were derived
from modeling the \water\ and SiO line brightness temperatures in the
large-velocity gradient (LVG) approach. We adopted the collisional coefficients
of Faure et al. (2007) and Dayou \& Balan\c{c}a (2006), respectively.
The gas density can be constrained reasonably well from the ratio of the
SiO \jeighttoseven\ to
\jthreetotwo\ brightness temperatures. This ratio was computed using the SiO
\jthreetotwo\ observations in the Cep\,E protostellar core obtained by L96 (see
also Fig.~3). We note that the source is unresolved by the single-dish JCMT and
IRAM 30m observations; therefore, we approximated the ratio of the line
brightness temperatures by the ratio of the main-beam brightness temperatures
corrected for the difference of main-beam solid angles $\rm R(8-7/3-2)$. A
description of the radiative transfer code and the procedure applied to deal
with the masing effect is presented in Cernicharo et al. (1994) and
Gonz\'alez-Alfonso \& Cernicharo (1993).

The top right panel of Fig.~3 displays the variations of $\rm R(8-7/3-2)$. The
ratio varies little  in the low-velocity gas and is about 0.2. It increases as
a function of velocity and reaches values up to 0.7 and 1.5 in the redshifted
and blueshifted gas, respectively. The variations of $\rm R(8-7/3-2)$ as a
function of $\rm T_{k}$ and $\rm n(\htwo)$ were computed for a typical column
density $\rm N(SiO)= 2\times 10^{14}\cmmd$  and a linewidth $\Delta V= 10\kms$, and
they are displayed in Fig.~3. With the
constraint $T_k \geq 100\K$, it comes out that $n(\htwo)$  must be about
$1.0\times 10^6\cmmt$ in order to account for the line intensity ratio in the
redshifted gas . The density in the blueshifted gas is slightly higher  $\approx
2\times 10^6\cmmt$. These calculations show that the ratio depends weakly on
$T_k$ above $100\K$. For these values of  $n(\htwo)$ and the kinetic temperature estimated
above, we estimate the  source-averaged column density necessary to account for
the \jeighttoseven\ line intensity $\rm N(SiO)= (2.0-3.3)\times 10^{14}\cmmd$ for gas at
$120\K$ in the blue- and redshifted components.
Under such physical conditions both SiO transitions are optically thin, so that
R(8-7/3-2) depends weakly on the adopted value of the gas column
density. From comparison with N(CO) and assuming a standard abundance $\rm [CO]/[\htwo]=
10^{-4}$, we obtain the relative SiO abundance in the shock~: [SiO]/[$\htwo] \simeq 1.0\times
10^{-7}$, similar to  the abundances measured in other Class~0 protostellar
outflows (Lefloch et al. 1998).

\setcounter{figure}{3}
\begin{figure}
\includegraphics[height=0.9\columnwidth,angle=-90]{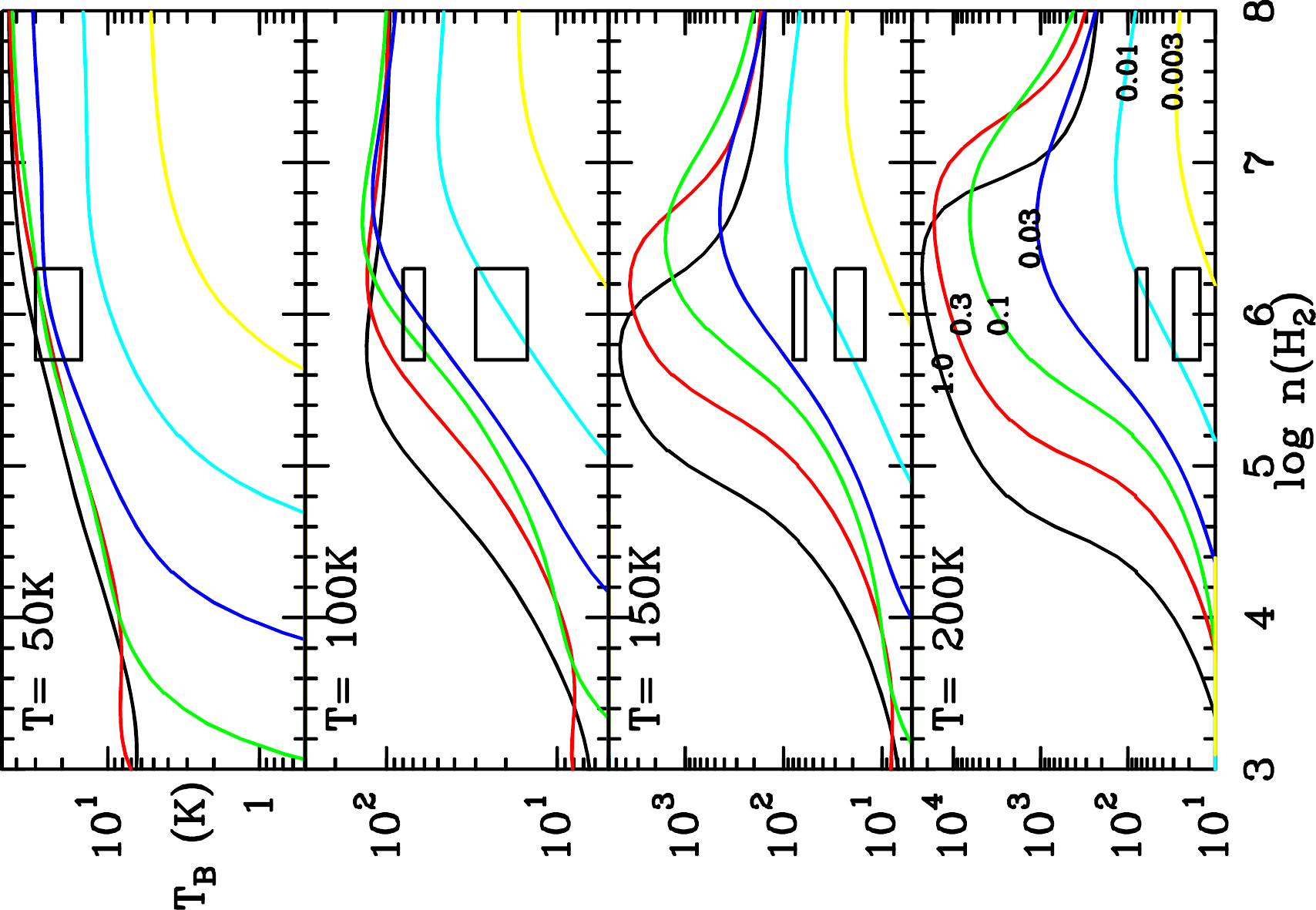}
\caption{Predictions of the \pwater\ line intensity, in units of brightness
temperature, for gas with density $n(\htwo)$ in the range $10^3 - 10^8\cmmt$,
temperature in the range $50 - 200\K$, for column densities of para-water equal
to 0.003 (bottom curve), 0.01, 0.03, 0.1, 0.3, $1.0\times 10^{19}\cmmd$ (top
curve). The range of $\rm T_B - n(\htwo)$ determined in the water-emitting
clumps is delineated by rectangles.}
\end{figure}

The para-water (source-averaged) column density $\rm N(p-\water)$ was determined from
modeling the line brightness temperature $\rm T_B$. The latter was derived
from the main-beam brightness temperature $T_{MB}= T_{B}\times ff$, where the filling
factor $ff$ was obtained assuming that the $\water$ and SiO \jtwotoone\
emitting regions have similar sizes, in agreement with Sect.~3.1. We find $T_B=
60-80\K$ and $15-29\K$ in the blueshifted  and redshifted gas, respectively.
The brightness temperature of the \pwater\ line was computed for a wide range
of physical conditions, adopting a typical linewidth $\Delta v= 10\kms$~: $\rm
T_k$ between 50 and $200\K$, $\rm n(\htwo)$ in the range $10^3 - 10^8\cmmt$ and
$\rm N(p-\water)$ in the range $3\times 10^{16}$ to $10^{19}\cmmd$. The results
of the calculations are presented in Fig.~4.

It comes out that the kinetic temperature $\rm T_k$ has to be higher than
$100\K$   to account for brightness temperatures of $60-80\K$. Our
calculations indicate a p-\water\ column density in the range $(0.3-2.0)\times
10^{17}\cmmd$ for $\rm T_k \simeq 200\K$. Assuming an ortho-to-para
ratio of 3, we derive $\rm N(\water)= (1.2-8.0)\times 10^{17}\cmmd$, and using
the \htwo\ column density derived in the high-velocity clumps, we obtain the
water abundance $\rm X(\water)= (0.5-3.2)\times 10^{-4}$. In this range of physical conditions,
the line optical depth $\tau$ is small $\approx -1$ and shows evidence for the
weak masing amplification of \pwater\ line. Suprathermal excitation of the line
would require much higher \water\ column densities. This is
possible if the emission arises from a small region of subarsec size
(see e.g. Cernicharo et al. 1994) and is very sharp. The line $\rm T_B$
would then look  weaker as a result of the dilution in the $2\kms$ channels
of the spectrometer.

Moro-Martin et al. (2001) analyzed the far-infrared emission lines of CO and
\water\ in Cep\,E, as observed with ISO.  Like Giannini et al. (2001), they
found that the line emission can be accounted for by an extended component of
low-density gas ($2\times 10^4\cmmt$) at $\sim 1000\K$. They show that the CO
and \water\ lines could be reproduced equally well by "warm" gas at a much
lower temperature ($\sim 200\K$) and higher density ($2\times 10^6\cmmt$). The
physical conditions we derive for the \pwater\ emitting region are actually in
rather good agreement with this "warm" solution and inconsistent with the
low-density "hot" solution.

\subsection{Shocked water in Cep\,E}

Crimier et al. (2010) determined the physical conditions in the protostellar
envelope of Cep\,E envelope and found $\rm n(\htwo)= 1.0\times 10^7\cmmt$ and
$T_k \sim 50\K$, respectively, at the location of the water-emitting clumps, i.e.
1000~AU ($1.4\arcsec$) from the protostar. No \pwater\ emission is detected
from the envelope at the
$2\sigma$ level $= 3\K$ ($\rm T_B$). From Fig.~4, we conclude that $N(\water) <
10^{16}\cmmd$ in the inner protostellar envelope, and the water abundance is
enhanced by 1 to 2 orders of magnitude in the high-velocity clumps. The authors
estimate a radius of $\approx 220~AU (0.3\arcsec)$ for the hot
corino region, and the density is predicted very high $\sim 10^8\cmmt$; again,
no masing effect is to be expected, in agreement with our observations.

The water-emitting clumps are located along the high-velocity jet, which
suggests that they are closely related with the jet (Fig.~2). One possibility is that
these clumps are tracing the jet/envelope interaction. However, the density of
the protostellar envelope, as estimated by Crimier et al.
(2010) at the location of the water emitting-clumps ($n(\htwo)= 1.0\times 10^7\cmmt$),
is about one order of magnitude higher
than the density measured in the shocked gas ($10^6\cmmt$), in apparent contradiction with this
hypothesis. The structure of the envelope could be actually more complex than the simple power-law distribution derived by Crimier et al. (2010), as  proposed that multiple
protostars are embedded inside the core (Ladd \& Hodapp, 1997). Direct observational
evidence for several condensations is still missing, however. Alternatively,  we speculate that
the water emission could arise from internal shocks in the entrained material of the flow.
 Interferometric observations at
subarcsec scale  in the inner core ($r < 1000$~AU) are needed to determine the gas density and velocity field in order to  determine the location and the origin of the water-emitting shocks, as well as their
relation with the jet.

Simple calculations in the LVG approximation  predict that the \water\ emission
from the high-velocity clumps should be easily detected in several
ortho- and para-\water\ transitions with the high-resolution spectrometer HIFI
onboard {\em Herschel}.
Such observations will allow us to better constrain the shock physical conditions
(see e.g. Flower \& Pineau des Forets, 2010).

\acknowledgements We thank an anonymous referee and the Editor for comments that
helped to improve the manuscript a lot. This project was supported by the research grant SAB2009-0011
of the Spanish Ministry of Education.

\def\baselinestretch{0.8}

{}

%
% figure 3
\setcounter{figure}{2}
\begin{figure}
\includegraphics[width=\columnwidth]{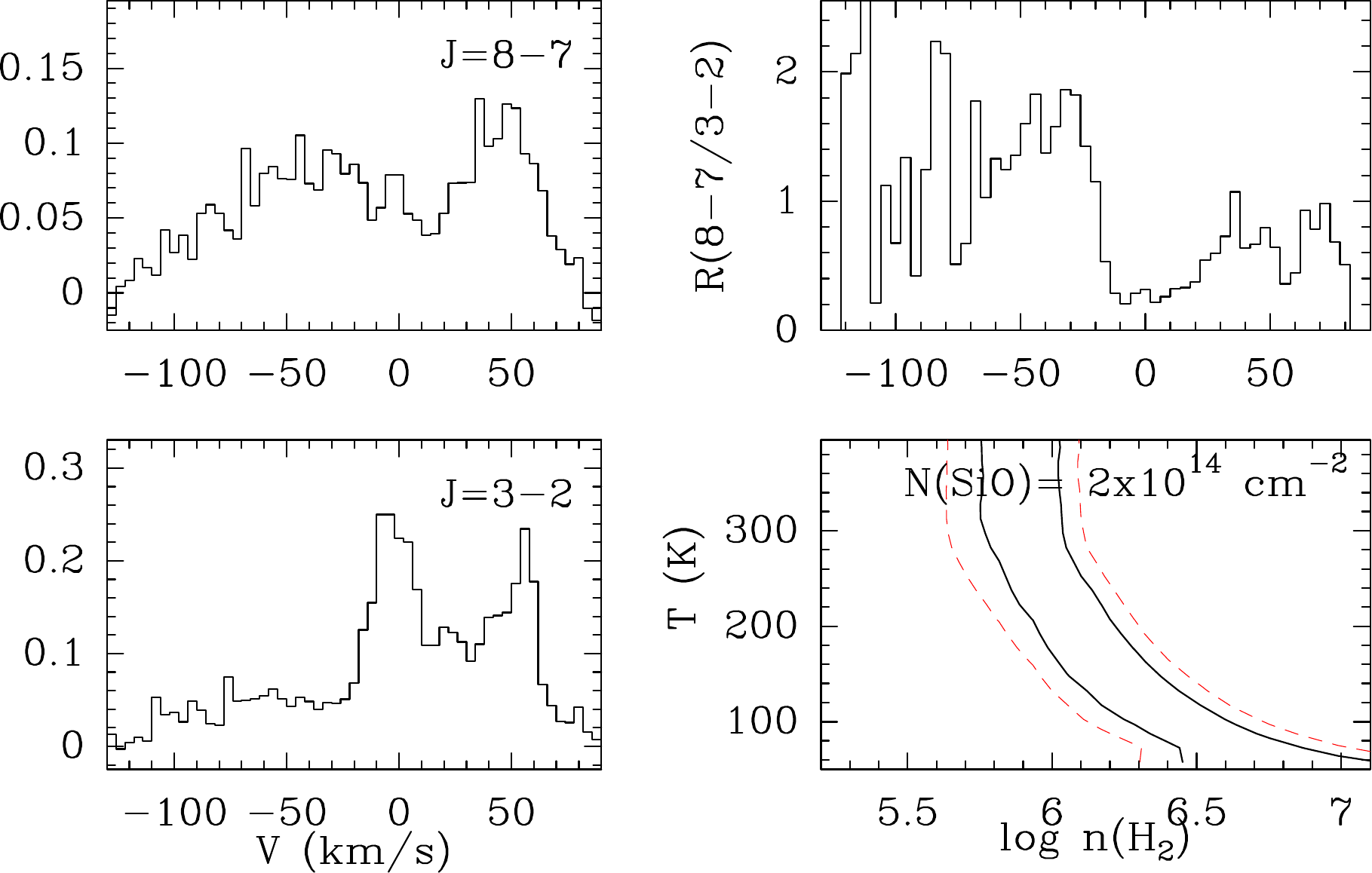}
\caption{{\em (left)}~SiO emission observed in the \jthreetotwo\ (bottom) and
\jeighttoseven\ transitions towards the protostellar source. Fluxes are
expressed in units of main-beam temperature. {\em (right)}~Observed variations
of the ratio of the CO $\jeighttoseven$ to $\jthreetotwo$ brigthness
temperatures R(8-7/3-2) as a function of velocity (top) and predicted
variations as a function of the gas density $n(\htwo)$ and the kinetic
temperature $T_k$ (bottom) for a typical column density $\rm N(SiO)=
2\times 10^{14}\cmmd$. The contour level R= 0.7, 1.5 are drawn with a solid black line; the
contours R= 0.5, 1.8 are drawn with a dashed red line and account for an
uncertainty of $30\%$.
}
\end{figure}

\end{document}